\documentclass[pdftex,11pt]{article}
\usepackage{url}
\usepackage{graphicx}
\usepackage[usenames]{color}
\usepackage{subfigure}
\usepackage{latexsym}
\usepackage{hyperref}
\usepackage{pslatex}
\definecolor{DodgerBlue4}{rgb}{0.06,0.31,0.55}

\hypersetup{
  pdftitle={Algorithmic Based Fault Tolerance Applied to High Performance Computing},
  pdfauthor={Julien Langou}, 
  linkbordercolor={0 0 0},
  citebordercolor={0 0 0},
  urlbordercolor={0 0 0},
  pdfborder={0 0 0},
  colorlinks={true},
  linkcolor=DodgerBlue4,
  urlcolor=none,
  citecolor=DodgerBlue4
}

\begin{document}

\noindent{\color{DodgerBlue4}\Large \textbf{Algorithmic Based Fault Tolerance Applied to High Performance Computing}}\\

\noindent \textbf{\color{DodgerBlue4} May 23, 2008}\\

\noindent
\textbf{\color{DodgerBlue4}George Bosilca}\\
Department of Electrical Engineering and Computer Science, University of Tennessee\\
\textbf{\color{DodgerBlue4}R\'emi Delmas}\\
Department of Electrical Engineering and Computer Science, University of Tennessee\\
\textbf{\color{DodgerBlue4}Jack Dongarra}\\
Department of Electrical Engineering and Computer Science, University of Tennessee\\
\textbf{\color{DodgerBlue4}Julien Langou}\\
Department of Mathematical and Statistical Sciences, University of Colorado Denver\\

\begin{center}
\begin{minipage}{14cm}
\noindent\textbf{\color{DodgerBlue4}Abstract:} {\small
We present a new approach to fault tolerance for High Performance Computing
system.  Our approach is based on a careful adaptation of the Algorithmic Based
Fault Tolerance technique (Huang and Abraham, 1984) to the need of parallel
distributed computation.  We obtain a strongly scalable mechanism
for fault tolerance.  We can also detect and correct errors (bit-flip) on the fly
of a computation.  To assess the viability of our approach, we have developed
a fault tolerant matrix-matrix multiplication subroutine and we propose some
models to predict its running time.  Our parallel fault-tolerant
matrix-matrix multiplication scores 1.4 TFLOPS on 484 processors (cluster
\texttt{jacquard.nersc.gov}) and returns a correct result while one process
failure has happened. This represents 65\% of the machine peak efficiency and
less than 12\% overhead with respect to the fastest failure-free
implementation. We predict (and have observed) that, as we increase the
processor count, the overhead of the fault tolerance drops significantly.
}
\end{minipage}
\end{center}

\section*{\color{DodgerBlue4}1~~~~Introduction}

Much research has been conducted into the checkpointing of message
passing applications~\cite{BLKC:03}. The earlier project proposing
checkpoint/restart facilities for a parallel application based on MPI
was CoCheck~\cite{Stel:96} and some of the most current and widely
used systems are LAM/MPI~\cite{BuDV:94}, MPICH-V/V2/V3~\cite{BLKC:03},
CHARM++~\cite{CHARM++} and Open MPI~\cite{ompiv:08}. The foundation
for all of these projects is that the system does not force the MPI
application developers to handle the failures themselves, i.e., the
underlying system, be it the Operating System or the MPI library
itself, is responsible for failure detection and application recovery.
The user of the system is not directly burdened with this at the MPI
API layer.

System-level checkpointing in parallel and distributed computing
settings has been studied extensively.  The issue of coordinating
checkpoints to define a consistent recovery line has received the
attention of scores of papers, ably summarized in the survey paper of
Elnozahy et al~\cite{EAWJ:02}.  Nearly all implementations of
checkpointing (e.g.~\cite{AgFr:99, Casa:95, ClGi:96, ElJZ:92, LeFS:93,
  LiTs:97, NaMM:97, PrLi:96, SiSi:92, Stel:94, Stel:96}) are based on
globally coordinated checkpoints, stored to a shared stable storage.
The reason is that system-level checkpointers are complex pieces of
code, and therefore real implementations typically keep the
synchronization among checkpoints and processors
simple~\cite{HuWa:95}.  While there are various techniques to improve
the performance of checkpointing (again, see \cite{EAWJ:02} for a
summary), it is widely agreed upon that the overhead of writing data
to stable storage is the dominant cost.

Diskless Checkpointing has been studied in various guises by a few
researchers.  As early as 1994, Silva et al explored the performance
gains of storing complete checkpoints in the memories of remote
processors~\cite{SIVS:94}.  Kim et al. \cite{PlKD:95} presented a
similar idea of diskless checkpointing in 1994 as well.  This
technique was subsequently revised for SIMD machines by
Chiueh~\cite{ChDe:96}.  More recently, diskless checkpointing of FFTs
has been studied in~\cite{EnGe:03}.  Evaluations of diskless and
diskless/disk-based hybrid systems have been performed by
Vaidya~\cite{Vaid:95}.  Lu presents a comparison between diskless
checkpointing and disk based checkpointing~\cite{Lu:2005}.

While these techniques can be effective in some specific cases,
overall, automatic \textit{application--oblivious} checkpointing of
message passing applications do suffer from scaling issues and in some
cases can incur considerable message passing performance penalties.

A relevant set-up of experimental conditions considers a constant
failure rate per processor. Therefore, as the number of processors
increases, the overall reliability of the system decreases
accordingly.  Elnozahy and Plank~\cite{ElPl:04} proved that, in these
conditions, checkpointing-restart is not a viable solution.  Since the
failure rate of the system is increasing with the number of
processors, a scalable application requires its recovery time to
decrease as the number of processors increases. The checkpoint-restart
mechanism does not enjoy this property (at best the cost for recovery
is constant).

Our contribution with respect to existing HPC fault tolerant research
is to present a methodology to enrich existing linear algebra kernels
with fault tolerance capacity at a low computational cost. We believe
it is the first time that a technique is devised that enables a
fault-tolerant application to be able to reduce the fault tolerance
overhead while the number of processors increases and the problem size
is kept constant. Moreover not only can our method recover from
process failures, it can also detect, locate and correct ``bit-flip''
errors in the output data. The cause of these errors (communication
error, software error, etc.) is not relevant to the overall
detection/location/correction process.

\textit{Algorithm-based fault tolerance} (ABFT), originally developed
by Huang and Abraham~\cite{HuAb:84}, is a low-cost fault tolerance
scheme to detect and correct permanent and transient errors in certain
matrix operations on systolic arrays. The key idea of the ABFT
technique is to encode the data at a higher level using checksum
schemes and redesign algorithms to operate on the encoded data. These
techniques and the flurry of papers that augmented them
\cite{BaAb:86,Bane:90,LuPa:88,RoBa:94} opened the door for jettisoning
system-level techniques and instead focused on high-performance
fault-tolerance of matrix operations.



The present manuscript focuses on exposing a new technique specific to
linear algebra based on the ABFT approach from Huang and
Abraham~\cite{HuAb:84}.  Our contribution with respect to the original
work of Huang and Abraham is to extend ABFT to the parallel
distributed context. Huang and Abraham were concerned with error
detection, location and recovery in linear algebra operation. Once a
matrix-matrix multiplication is performed (for example), then ABFT
enables to recover from errors.  This scenario is not ideal for HPC
where we want to be able to recover an errorless environment
immediately after a failure.  The contribution is therefore to create
algorithms for which ABFT can be used on the fly.

This manuscript focuses on obtaining an efficient fault-tolerant
\textit{matrix-matrix multiplication subroutine} (PDGEMM).  This
application does not respond well to memory exclusion
techniques~\cite{KiPD:97}, therefore standard checkpointing techniques
perform poorly.  Matrix-matrix multiplication is a kernel of
fundamental importance to obtain efficient linear algebra subroutines.
Our claim is that we can encapsulate all the fault tolerance needed by
the linear algebra subroutines in ScaLAPACK in a fault-tolerant
\textit{Basic Linear Algebra Subroutines} (BLAS).

The third contribution of this manuscript is the presentation of model
to predict the running time of our routines.

The fourth contribution of this manuscript is a software based on
FTMPI and some experimental results.  Our parallel fault-tolerant
matrix-matrix multiplication scores 1.4 TFLOPS on 484 processors
(cluster \texttt{jacquard.nersc.gov}) and returns a correct result
while one process failure has happened. This represents 65\% of the
machine peak efficiency and less than 12\% overhead with respect to
the fastest failure-free implementation.

\section*{\color{DodgerBlue4}2~~~~A new approach for HPC fault
  tolerance}

\subsection*{\color{DodgerBlue4}2.1~~~~Additional processors to
  store redundant information}

During a computation, our data is spread across different
processes. If one of these processes fails, we need an efficient way
to recover the lost part of the data. In that respect, we use
additional processes to store redundant information in a checksum
format.  Checksums represent an efficient and memory--effective way of
supporting multiple failures.

If a vector of data $x$ is spread across $p$ processes where $x_i$ is
held by process $i$, then an additional process is added for the
storage of $y$ such that $ y = a_{1}x_1 + \ldots + a_{p} x_p $.  (For
the sake of simplicity, we assume the size of $x$ is constant on all
the processes). In case of a single process failure, using the
information in the additional checksum process and the non-failed
processes, the data on the failed process can trivially be restored.
(Assuming the $a_i$ are not 0.)  This fault-tolerant mechanism is
classically known as diskless checkpointing and was first introduced
by Plank et al.~\cite{PlKD:95}. The name diskless comes from the fact
that checksums are stored on additional processes as opposed to being
stored on disks.

In order to support $f$ failures, $f$ additional processes are added
and $f$ checksums are performed with the following linear relation
$$
\left\{\begin{array}{l}
 y_1 = a_{11}x_1 + \ldots + a_{1p} x_p,\\
\vdots \\
 y_f = a_{f1}x_1 + \ldots + a_{fp} x_p.
\end{array}\right. 
$$
A sufficient condition to recover from any $f$-failure set is that any
$f$--by--$f$ submatrix of the $f$--by--$p$ matrix $A$ is nonsingular.

Checksums are traditionally performed in Galois Field
arithmetic. Another natural choice to encode floating-point numbers is
to use the floating-point arithmetic. Galois Field always guarantees
bit-by-bit accuracy. Floating-point arithmetic suffers from numerical
errors during the encoding and the recovery. However, cancellation
errors in the checksum are typically of the same order as the ones
arising in the numerical methods, therefore of no concern for the
quality of the final solution; additionally, the $f$--by--$f$ recovery
submatrix is statistically guaranteed to be well-conditioned if we
take the checkpoint matrix $A$ with random values~\cite{ChDo:05}.

The ABFT technique presented in this manuscript is based on a
floating-point checksum.

\subsection*{\color{DodgerBlue4}2.2~~~~ABFT approaches to fault
  tolerance in FT-LA}

In 1984, Huang and Abraham~\cite{HuAb:84} proposed ABFT (Algorithm
Based Fault Tolerance) to handle errors in numerical computation. This
section relies heavily on their idea. To illustrate the ABFT
technique, we consider two vectors $x$ and $y$, spread among $p$
processes, where process $i$ holds $x_i$ and $y_i$. An additional
checksum process has been added to store the checksum $x_c = x_1 +
\ldots + x_p $ and the checksum $y_c= y_1 + \ldots + y_p$. The
checksums are performed in floating-point arithmetic. Assume now that
one wants to realize $z = x + y$.  This involves the local computation
on process $i$, $i=1,\ldots,p$ : $z_i = x_i + y_i$ and the update of
the checksum $z_c$. Instead of computing $z_c$ as $z_1 + \ldots +
z_p$, as proposed in Section~\color{DodgerBlue4}2.1\color{black},
ABFT simply performs $z_c= x_c + y_c$. While the traditional
checkpoint method would require a global communication among the $n$
processes and an additional computational step, ABFT performs a local
operation (no communication involved) on an additional process during
the time the active processes perform the same kind of operation.
Therefore, to maintain the checksum of $z_c$ consistent with the
vector $z$, the penalty cost with respect to the non-fault-tolerant
case is an additional process.

The same idea applies to all linear algebra operations. For example
matrix-matrix multiplication, LU factorization, Cholesky factorization
or QR factorization (see
\cite{BaAb:86,Bane:90,HuAb:84,LuPa:88,RoBa:94}).

\begin{figure}[!h]
\color{DodgerBlue4}
\begin{center}
\includegraphics[width=0.50 \textwidth]{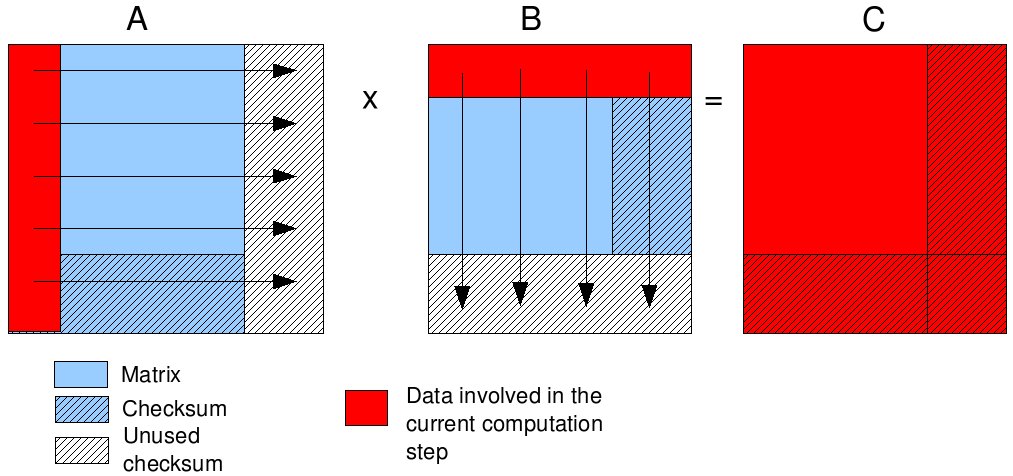}
\end{center}
\caption{\label{fig:dgemm}
\color{black}
Algorithm-Based Fault-Tolerant DGEMM
}
\end{figure}

As an example, we have studied ABFT in the context of the
matrix-matrix multiplication.  Assuming that $A$ and $B$ have been
checkpointed such that
$$
   A_F = \left(\begin{array}{cc} A & AC_R \\ C_C^TA & C_C^TAC_R \end{array} \right) 
\quad \textmd{and} \quad
   B_F = \left(\begin{array}{cc} B & BC_R \\ C_C^TB & C_C^TBC_R \end{array} \right) ,
$$
where $C_C$ and $C_R$ are the checksum matrices; then performing the
matrix-matrix multiplication
\begin{equation}\label{eq:ABFT-GEMM}
   \left(\begin{array}{cc} A \\ C_C^TA \end{array} \right) 
   \left(\begin{array}{cc} B & BC_R \end{array} \right) 
=
   \left(\begin{array}{cc} AB & ABC_R \\ C_C^TAB & C_C^TABC_R \end{array} \right) 
=  (AB)_F
,
\end{equation}
we obtain the result $(AB)_F$ which is \textit{consistent}, that is to
say, it verifies the same checkpoint relation as $A_F$ or $B_F$. This
consistency enables us to detect, localize and correct errors.  In the
context of erasure, one has to be a more careful since we want not
only the input and output matrices to be consistent, but we also want
any intermediate quantity to be consistent. Using the outer product
version of the matrix-matrix product (\texttt{for k=1:n,} $C_k = C_k +
A_{:,k}B_{k,:}$ \texttt{; end}), as $C_k$ is updated along the loop in
$k$, the intermediate $C_k$ matrices are maintained
consistently. Therefore a failure at any time during the algorithm can
be recovered (see Figure~\ref{fig:dgemm}).
 
If $p^2$ processes are performing a matrix-matrix multiplication,
maintaining the checksum consistent requires $(2p-1)$ extra processes.
The cost with respect to a non-failure-tolerant application occurs
when the information of the matrix $A$ is broadcasted along the
process rows. In this case, one extra processor needs to receive the
data. And vice versa for the matrix $B$: when the information of the
matrix $B$ is broadcasted along the process columns, one extra
processor needs to receive the data. These are the only extra costs
for being fault tolerant.
From this analysis, we deduce that the main cost to enable ABFT in a
matrix-matrix multiplication is to dedicate $(2p-1)$ processes over
$p^2$ for fault-tolerance sake. Therefore, the more processors, the
more advantageous the ABFT scheme is!

\section*{\color{DodgerBlue4}3~~~~Model}

\subsection*{\color{DodgerBlue4}3.1~~~~PBLAS PDGEMM}

It is behind the scope of this manuscript to describe every aspect of
parallel matrix-matrix multiplication, and we refer the reader to
\cite{GeWa:97} for more information.

During a matrix-matrix multiplication, each block of $A$ in the
current column (of size $mloc*nb$) is broadcasted along the row to all
other processes. The same happens to $B$: each block in the current
row (of size $nb*nloc$) is broadcasted along the columns. After these
two broadcasts, each process on the grid computes a local
\textsc{dgemm}. This step is repeated until the whole $A$ and $B$
matrices have been broadcasted to every process.

For the sake of simplicity, we consider that the matrices are square,
of size $n$--by--$n$, and distributed over a square grid of
$p$--by--$p$ processes.

In SUMMA, we implement the broadcast as passing a message around the
logical ring that forms the row or column. In that case the time
complexity becomes

\begin{equation}
(\sqrt{p}-1)(\alpha + \frac{n}{\sqrt(p)}\beta)+(\sqrt{p}-1)(\alpha+\frac{n}{\sqrt{p}}\beta)
\label{eq:summa1}
\end{equation}

\begin{equation}
~~~~~~+k(\frac{2n^{2}}{p}\gamma+\alpha+\frac{n}{\sqrt{p}}\beta+\alpha+\frac{n}{\sqrt{p}}\beta)
\label{eq:summa2}
\end{equation}

\begin{equation}
~~~~~~+(c-2)(\alpha+\frac{n}{\sqrt{p}}\beta)+(\sqrt{p}-2)(\alpha+\frac{n}{\sqrt{p}}\beta)
\label{eq:summa3}
\end{equation}

\begin{equation}
~~~~~~+\frac{2n^{3}}{p}\gamma
\label{eq:summa4}
\end{equation}

\begin{equation}
= \frac{2n^{2}(n+1)}{p}\gamma+ 2(n+2\sqrt{p}-3)(\alpha+\frac{n}{\sqrt{p}}\beta)
\label{eq:summa5}
\end{equation}

where $\alpha$ is the inverse of the bandwidth, $\beta$ is the latency
and $\gamma$ is the inverse of the flop rate.

Equation~(\ref{eq:summa1}) is the time required to fill the pipe (that
is the time for the messages originating from $A$ and $B$ to reach the
last process in the row and in the column). The next
term~(\ref{eq:summa2}) is the time to perform the sequential
matrix-matrix multiplication and passing the messages. Then,
contribution~(\ref{eq:summa3}) is the time for the final messages to
reach the end of the pipe. The last term is the time for the final
update at the node at the end of the pipe.

This complexity is then approximately

\begin{equation}
\frac{2n^{3}}{p}\gamma+2(n+2\sqrt{p}-3)(\alpha+\frac{n}{\sqrt{p}}\beta)
\end{equation}

and the estimated efficiency is

\begin{equation}
E(n,p)=\frac{1}{1+O(\frac{p}{n^{2}})+O(\frac{\sqrt{p}}{n})}
\end{equation}

We can see that the method is \textit{weakly} scalable: if we increase
the number of processors while maintaining constant the memory use per
node (thus having $\frac{p}{n^{2}}$ constant), this algorithm
maintains its efficiency constant.

In the remainder, we neglect the latency term ($\beta$) in the
communication cost.

\subsection*{\color{DodgerBlue4}3.2~~~~ABFT PDGEMM (0 failure)}

In this section, we derive a model for ABFT PDGEMM.  If we perform a
traditional matrix-matrix multiplication, the result will be a
checkpointed matrix (see Equation~\ref{eq:ABFT-GEMM}). In the
outer-product variant of the matrix-matrix multiplication algorithm,
rank-$nb$ updates are applied to the global matrix $C$. Therefore, the
checksum stays consistent throughout the execution of the algorithm
provided that all the processes go at the same speed.

The last row of $A$ (checksum) and the last column of $B$ (checksum)
are sent exactly in the same way as the rest of the data (see
Figure~\ref{fig:dgemm}).

When no failure occurs, the overhead of the fault tolerance are:
\begin{itemize}

\item the initial checksum. However, if we call ABFT BLAS functions
  the ones after the other, we do not have to recompute the checksum
  between each call. As a consequence, we do not consider the cost of
  the initial checksum.

\item the computation of an $(n+nloc)$--by--$n$--by--$(n+nloc)$
  matrix--matrix multiplication, instead of $n$--by--$n$--by--$n$ for
  a non fault-tolerant code.

\item the broadcast which needs to be performed on $q+1$ processes
  along the rows and $p+1$ processes along the columns, instead of
  respectively $p$ and $q$ for a non fault-tolerant matrix-matrix
  multiplication.

\end{itemize}

Using Equation~(\ref{eq:summa5}), we can then write the complexity for
ABFT PDGEMM with no fault :

\begin{equation}
\frac{2(n+nloc)^{2}n}{p}\gamma+2(n+2\sqrt{p}-3)(\frac{n+nloc}{\sqrt{p}}\beta)
\label{eq:effsumma}
\end{equation}

The modification in the first term of the sum comes from the fact that
the matrix-matrix multiplication now involves $n$--by--$(n+nloc)$
matrices on the same number of processors. The modification in the
second term shows that the pipeline is now longer than in regular
SUMMA.

\subsection*{\color{DodgerBlue4}3.3~~~~ABFT PDGEMM (1 failure)}

\begin{figure}
\color{DodgerBlue4}
\begin{center}
\includegraphics[width=6cm]{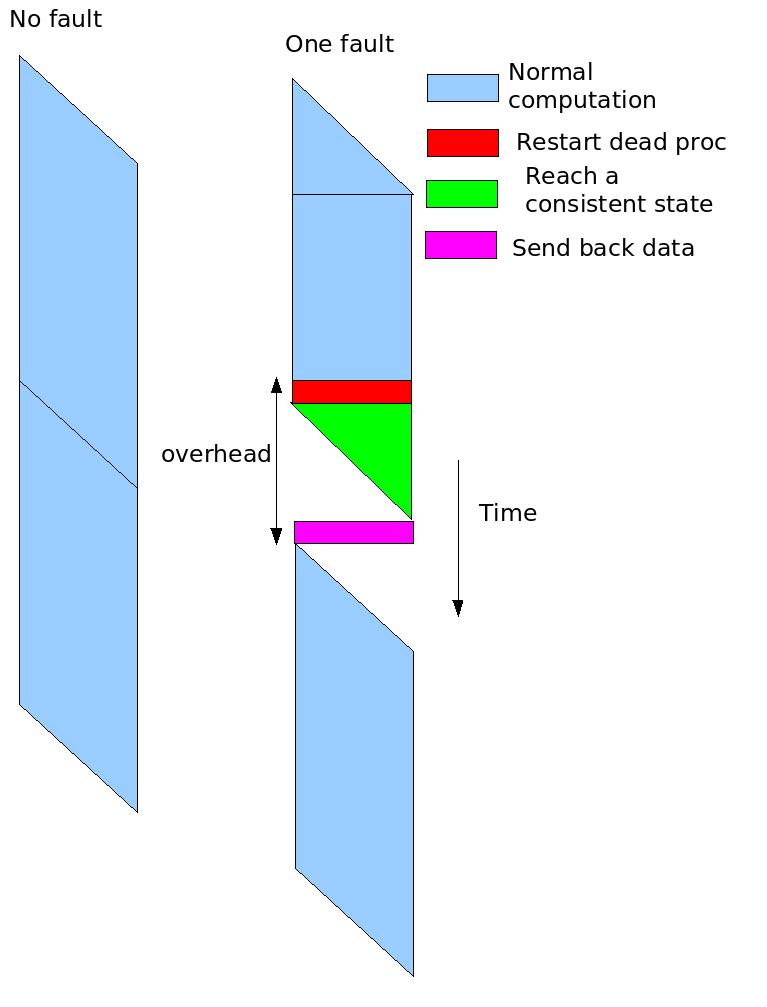}
\end{center}
\caption{\label{fig:recovery}
\color{black}
Overhead between no fault and one fault in ABFT SUMMA
}
\end{figure}

When one fault happens, the timeline is illustrated in
Figure~\ref{fig:recovery} and explained below.

\begin{itemize}

\item $T_{\textmd{\tiny{detection}}}$: All non-failed processes must
  be notified when a failure happens.  Since processes are only
  notified when attempting an MPI communication, odds are that at
  least one process will be doing a full local \textsc{dgemm} before
  being notified. So, basically, this overhead is more or less equal
  to the time of one local \textsc{dgemm}.

\item $T_{\textmd{\tiny{restart}}}$: FT-MPI must spawn a new process
  to replace the failed one. This is a blocking operation and the time
  for this step depends solely on the total number of processes
  involved in the computation.

\item $T_{\textmd{\tiny{pushdata}}}$: The non-failed processes must
  reach the same consistent state. The time overhead consists in
  filling and emptying the pipe once.

\item $T_{\textmd{\tiny{checksum}}}$: The last step in the recovery is
  to reconstruct the lost data of the failed process. This cost is the
  cost of an \textsc{mpi\_r}educe.

\end{itemize}

\section*{\color{DodgerBlue4}4~~~~Experimental results}

\subsection*{\color{DodgerBlue4} 4.1~~~~ABFT BLAS framework}

\begin{figure}
  \color{DodgerBlue4}
  \begin{center}
    \includegraphics[width=0.4\textwidth]{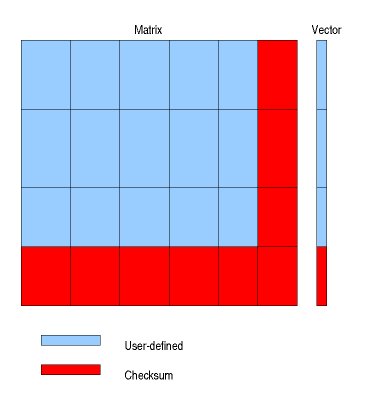}
    \caption{\label{fig:datadistribution} \color{black} Data distribution
      in the ABFT BLAS framework for vectors and matrices over a 4-by-6
      process grid }
  \end{center}
\end{figure}

A prototype framework ABFT BLAS has been designed and implemented.
The ABFT BLAS application relies on the FT-MPI library for handling
process failures at the middleware level. The ABFT BLAS
responsibilities is to (efficiently) recover the data lost during a
crash.

Vectors over processes are created and then destroyed. Several
operations over the vectors are possible: scalar product, norm, vector
addition, etc.  Vectors are registered to the fault tolerant
context. When a failure occurs, all data registered in the fault
tolerant context is recovered and the application is continued. The
default encoding mode is diskless checkpointing with floating-point
arithmetic but an option is to perform Galois Field encoding (although
this rules out ABFT).

The same functionality have been implemented for dense matrices. Dense
matrices are spread on the processor in the 2D-block cyclic format,
and checksums can be performed by row, by columns or both (see
Figure~\ref{fig:datadistribution}).

Regarding matrix and vector operations, available functionality are
matrix-vector products and several implementations of matrix-matrix
multiplications. One of the main features of the ABFT BLAS library is
that the user is able to stack fault-tolerant parallel routines the
ones on top of the others. The code looks like a sequential code but
the resulting application is parallel and fault-tolerant.

\subsection*{\color{DodgerBlue4}
4.2~~~~Experimental setup}

All runs where done on the \texttt{jacquard.nersc.gov} cluster, from
the National Energy Research Scientific Computing Center (NERSC). This
cluster is a 512-CPU Opteron cluster running a Linux operating
system. Each processor runs at a clock speed of 2.2GHz and has a
theoretical peak performance of 4.4 GFLOPS/sec. Measured peak (on a
3000--by--3000--by--3000 \textsc{dgemm} run) is 4.03 GFLOPS/sec. The
nodes are interconnected with a high-speed InfiniBand network. The
latency is 4.5 $\mu$sec, while the bandwidth is 620 MB/sec.

Below are more details on the experimental set-up:
\begin{itemize}
\item we have systematically use square processor grids.
\item the blocksize for the algorithm is always 64.
\item process failure were performed manually. We introduced an
  \textsc{exit} statement in the application for one arbitrary
  non-checksum process.  Although our application support failures at
  any point in the execution (see
  Section~\color{DodgerBlue4}4.3\color{black}), we find that having
  a constant failure point was the most practical and reproduceable
  approach for performance measurements.
\item when we present results of ABFT PDGEMM on a $p$--by--$p$ grid,
  this means that the total number of processors used is
  $p^2$. Therefore $(p-1)^2$ processors are used to process the data
  while $(2p-1)$ are used to checkpoint the data.
\item The performance model parameters are
  \begin{center}flops rate: 3.75 GFLOPS/sec\hspace{1cm}bandwidth: 52.5
    MBytes/sec.\end{center} These two machine-dependent parameters and
  our models (see Section~\color{DodgerBlue4}3\color{black}) are
  enough to predict the running time for PBLAS PDGEMM and ABFT BLAS
  PDGEMM (0 failure).  To model ABFT BLAS PDGEMM (1 failure), we also
  need the time for a checkpoint (\textsc{mpi\_r}educe) and the time
  for the FT-MPI library to recover from a failure (see
  Section~\color{DodgerBlue4}3.3\color{black}).
\end{itemize}

\subsection*{\color{DodgerBlue4} 4.3~~~~Stress test}

In order to assess the robustness of our library, we have designed the
following stress test.  We set up an infinite loop where, at each
loop, we initialize the data ($A$ and $B$), checkpoint the data,
perform a matrix-matrix multiplication $ C \leftarrow AB $, and check
the result with residual checking.

Residual checking consists in taking a random vector $x$ and checking
that $\| Cx - A(Bx) \| / \left( n \varepsilon \| C \| \| x \| \right)
$ is reasonably small. If this test passes, then we have high
confidence of the correctness of our matrix-matrix multiplication
routine.

During the execution, a process killer is activated.  This process
killer kills randomly in time and in the location any process in the
application.

Our application have successfully returned from tens of such failures.
This testing not only stresses the matrix-matrix multiplication
subroutine but it also stresses the whole ABFT BLAS library since
failure can occurs at any time.

\subsection*{\color{DodgerBlue4} 4.3~~~~Performance results}

\begin{figure}
  \color{DodgerBlue4}
  \begin{center}
    \includegraphics[width=0.7 \textwidth,height=3in]{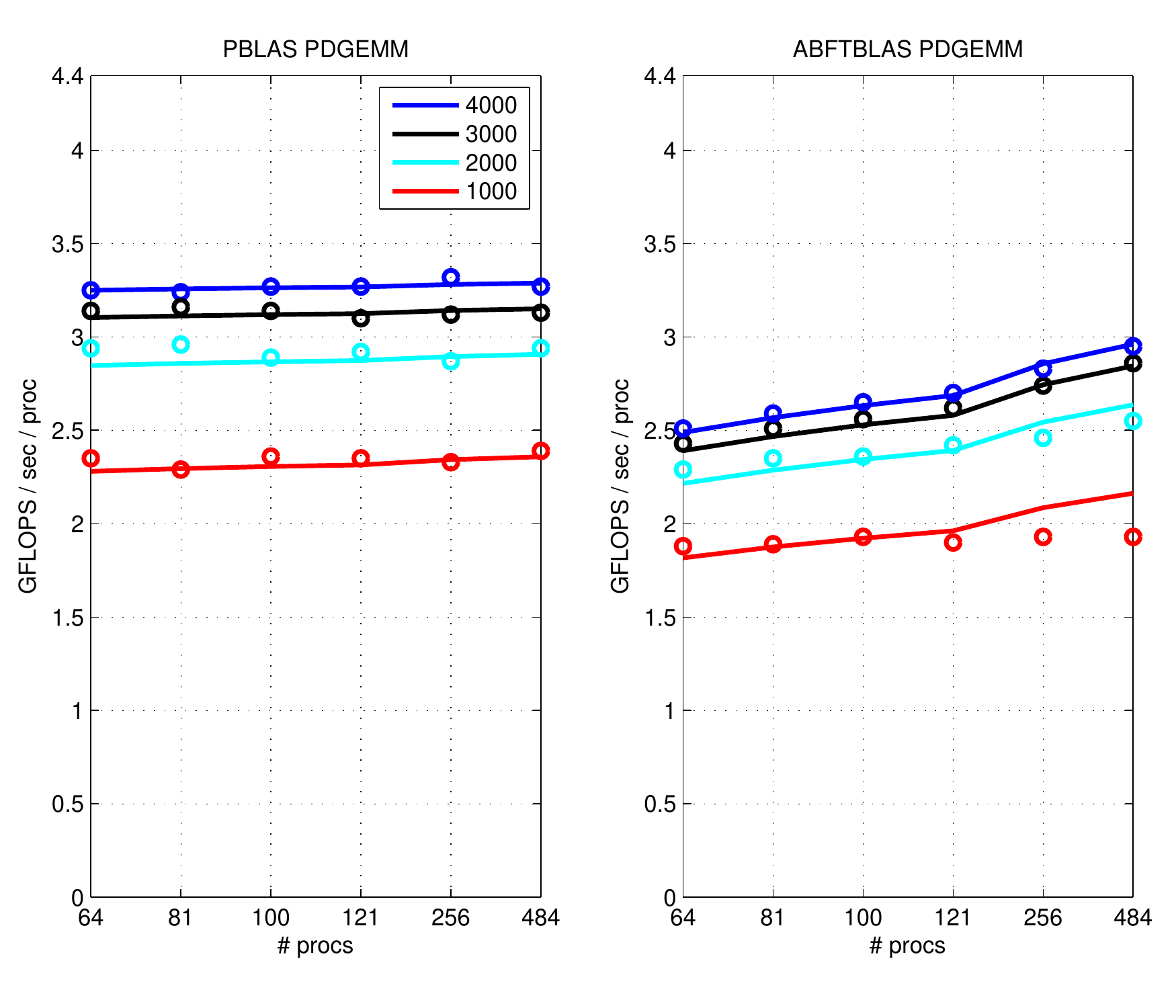}
    \color{DodgerBlue4}
    \caption{\label{fig:perf01} \color{black} Performance
      (GFLOPS/sec/proc) of PBLAS PDGEMM (left) and ABFT BLAS PDGEMM
      with 0 failure (right).  The solid lines represent model while
      the circles represent experimental points.  }
  \end{center}
\end{figure}

In Figure~\ref{fig:perf01}, we present a weak scalability experiment.
While the number of processors increases from 64 (8--by--8) to 424
(22--by--22), we keep the local matrix size constant ($nloc = 1000,
\ldots, 4000$) and observe the performance for PBLAS PDGEMM and ABFT
BLAS PDGEMM with no failure.

The first observation is that our model performs very well. With only
two parameters ($\alpha$ and $\gamma$), we are able to predict the 48
experimental values within a few percents.

We observe that PBLAS PDGEMM is weakly scalable. That is the
performance is constant when we increase the number of processors. We
also observe that as the processor count increases, the performance of
ABFT PDGEMM increases.

\begin{figure}
  \begin{center}
    \includegraphics[width=0.5 \textwidth]{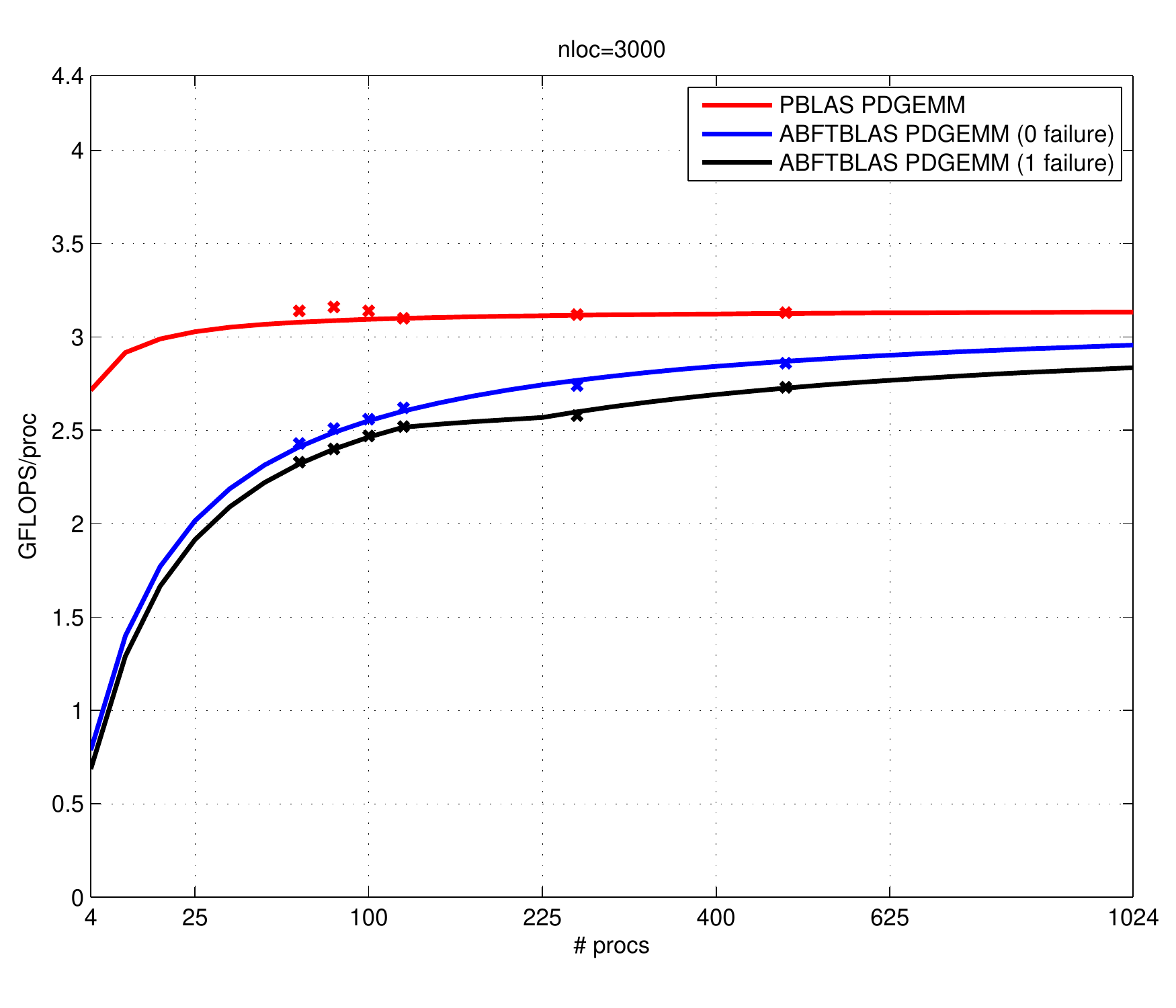}
  \end{center}
  \color{DodgerBlue4}
  \caption{\label{fig:perf02} \color{black} Performance
    (GFLOPS/sec/proc) of PBLAS PDGEMM, ABFT BLAS PDGEMM (0 failure),
    and ABFT BLAS PDGEMM (1 failure).  The solid lines represent model
    while the circles represent experimental points.  This is a weak
    scalability experiment with $nloc=3000$.  (See also
    Table~\ref{tab:perf02}.)  }
\end{figure}

\begin{figure}
  \begin{center}
    \includegraphics[width=0.5 \textwidth]{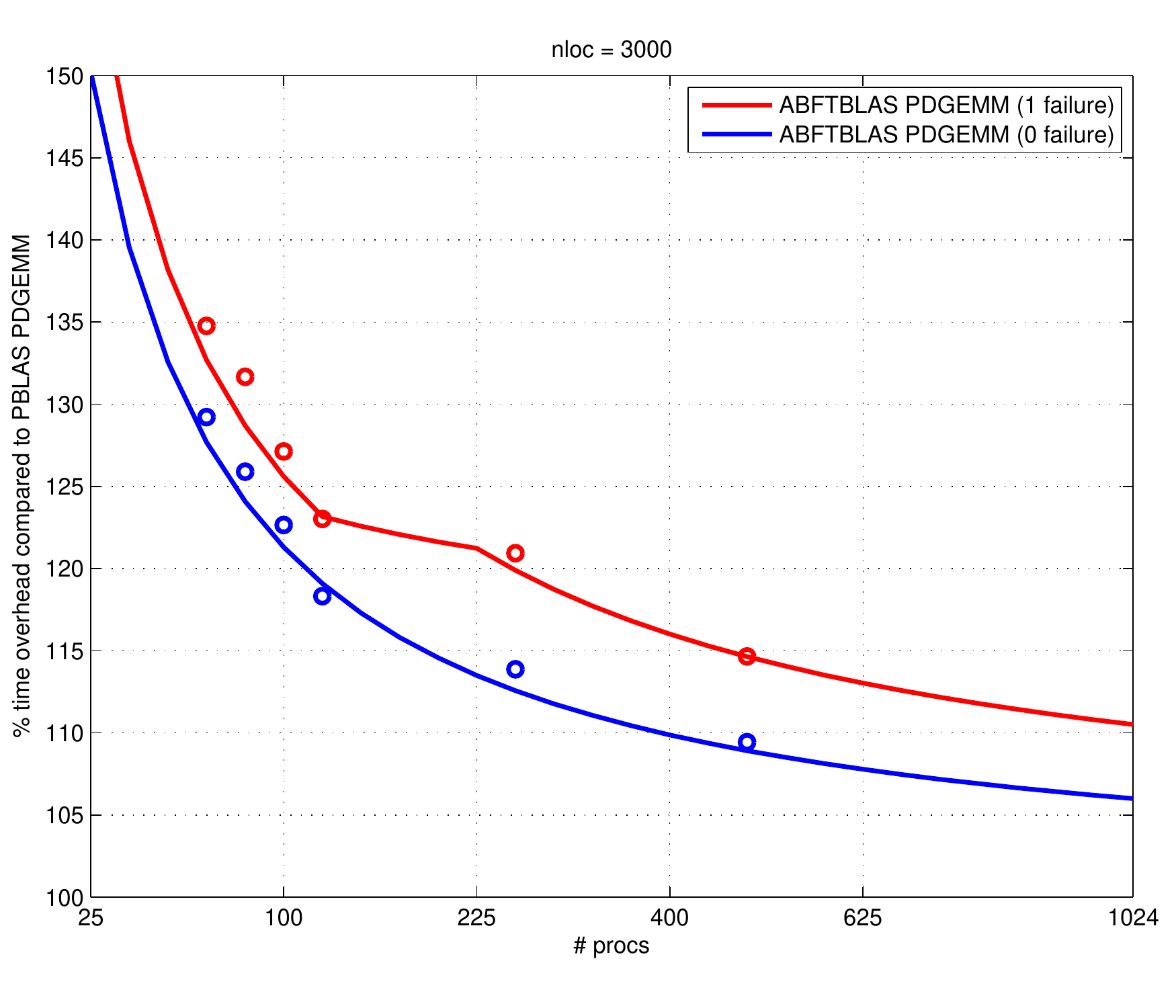}
  \end{center}
  \color{DodgerBlue4}
  \caption{\label{fig:perf03} \color{black} Overhead of the fault
    tolerance with respect to the non-failure-resilient application
    PBLAS PDGEMM.  The plain curves correspond to model while the
    circles correspond to experimental data.  This is a weak
    scalability experiment with $nloc=3000$.  (See also
    Table~\ref{tab:perf03}.)  }
\end{figure}

In Figure~\ref{fig:perf02}, Table~\ref{tab:perf02},
Figure~\ref{fig:perf03}, and Table~\ref{tab:perf03}, we present the
weak scalability of our application when $nloc$ is kept constant with
$nloc=3000$. We present performance results for PBLAS PDGEMM, ABFT
BLAS PDGEMM with 0 failure and ABFT BLAS PDGEMM with 1 failure.  We
see that as the number of processors increases the cost of the fault
tolerance converges to 0.

\begin{table}
  \begin{center}
    \small
    \begin{tabular}{|l||c|c|c|c|c|c|}
      \hline
      & \multicolumn{6}{|c|}{performance per processor (GFLOPS/sec/proc)}\\
      \hline
      &   64        &   81        &  100        &  121        &  256        &  484        \\
      \hline
      PBLAS PDGEMM                & 3.14 (3.09) & 3.16 (3.09) & 3.14 (3.10) & 3.10 (3.10) & 3.12 (3.12) & 3.13 (3.13) \\
      ABFTBLAS PDGEMM (0 failure) & 2.43 (2.49) & 2.51 (2.55) & 2.56 (2.60) & 2.62 (2.65) & 2.74 (2.79) & 2.86 (2.88) \\
      ABFTBLAS PDGEMM (1 failure) & 2.33 (2.40) & 2.40 (2.46) & 2.47 (2.52) & 2.52 (2.53) & 2.58 (2.63) & 2.73 (2.74) \\
      \hline
      & \multicolumn{6}{|c|}{cumul performance (GFLOPS/sec)}\\
      \hline
      &   64        &   81        &  100        &  121        &  256        &  484        \\
      \hline
      PBLAS PDGEMM                &  201 (198)  &  256 (251)  &  314 (310)  &  375 (376)  &  799 (798)  & 1515 (1513) \\
      ABFTBLAS PDGEMM (0 failure) &  156 (159)  &  203 (207)  &  256 (260)  &  317 (320)  &  701 (714)  & 1384 (1395) \\
      ABFTBLAS PDGEMM (1 failure) &  149 (154)  &  194 (200)  &  247 (252)  &  305 (306)  &  660 (672)  & 1321 (1327) \\
      \hline
    \end{tabular}
    \normalsize
  \end{center}
  \color{DodgerBlue4}
  \caption{\label{tab:perf02}
    \color{black}
    Performance (GFLOPS/sec/proc) of PBLAS PDGEMM, ABFT BLAS PDGEMM (0 failure), 
    and ABFT BLAS PDGEMM (1 failure). The number without parenthesis is the experimental result, 
    while the number in between parenthesis corresponds to the model value.
    This is a weak scalability experiment with $nloc=3000$.
    (See also Figure~\ref{fig:perf02}.)
  }
\end{table}

\begin{table}
  \begin{center}
    \begin{tabular}{|l||c|c|c|c|c|c|}
      \hline
      &   64        &   81        &  100        &  121        &  256        &  484        \\
      \hline
      PBLAS PDGEMM                &  100.0      &  100.0      &  100.0      &  100.0      &  100.0      &  100.0      \\
      ABFTBLAS PDGEMM (0 failure) &  129.2      &  125.9      &  122.7      &  118.3      &  113.9      &  109.4      \\
      ABFTBLAS PDGEMM (1 failure) &  134.8      &  131.7      &  127.1      &  123.0      &  120.9      &  114.7      \\
      \hline
    \end{tabular}
  \end{center}
  \color{DodgerBlue4}
  \caption{\label{tab:perf03}
    \color{black}
    Overhead of the fault tolerance with respect to the non-failure-resilient application PBLAS PDGEMM.
    This is a weak scalability experiment with $nloc=3000$.
    (See also Figure~\ref{fig:perf03}.)
  }
\end{table}

The fact that matrix-matrix multiplication performs $n^3$ operations
enables us to hide a lot of $n^2$ operations (for example
checkpointing) in the background.  This renders weak scalability
easily feasible.

\begin{figure}
  \begin{center}
    \mbox{
      \subfigure[Performance (GFLOPS/sec/proc)]
      {\includegraphics[width=0.5\textwidth]{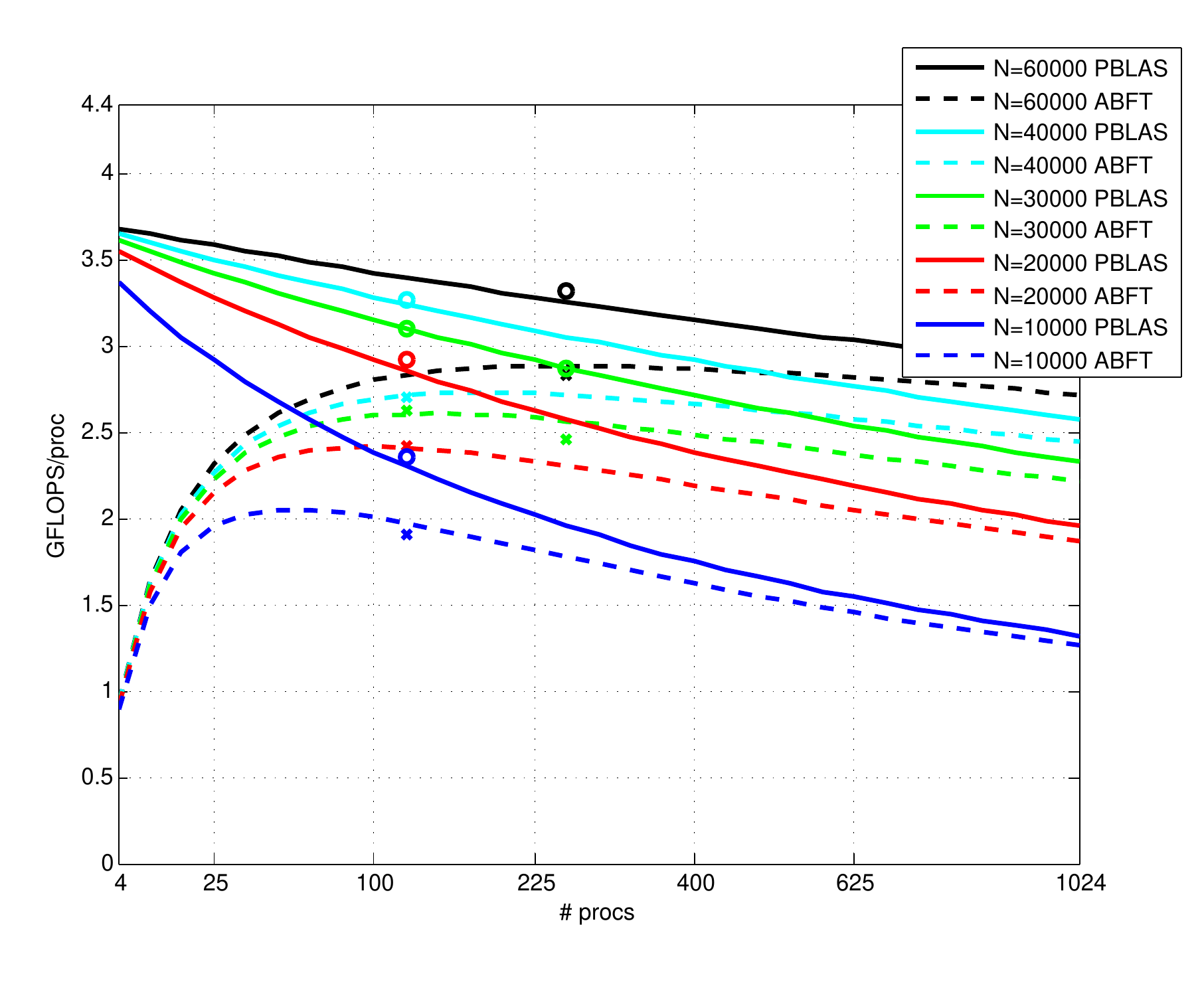}}
      \subfigure[Overhead with respect to PBLAS PDGEMM]
      {\includegraphics[width=0.5\textwidth]{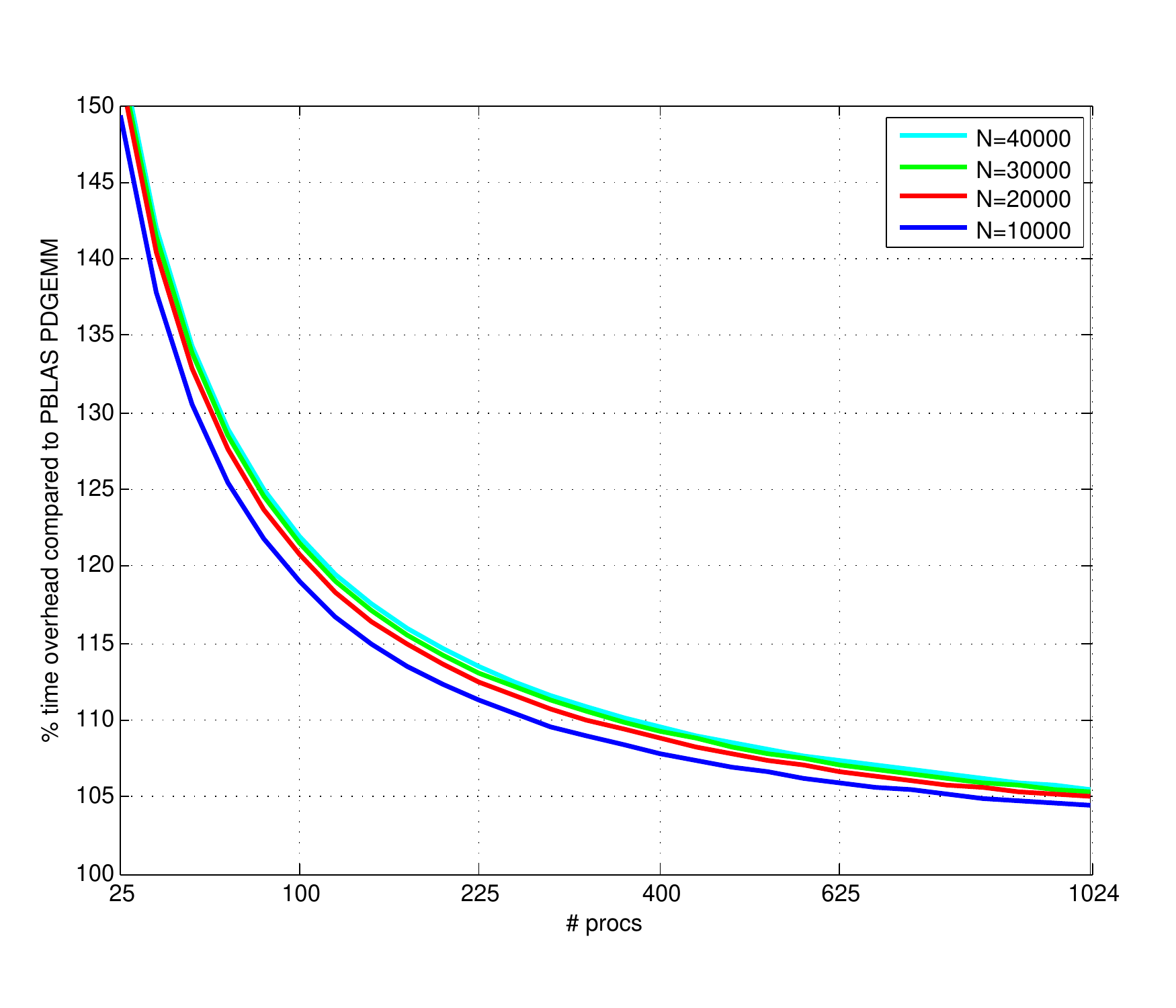}}
    }
  \end{center}
  \color{DodgerBlue4}
  \caption{\label{fig:perf04} \color{black} Strong scalability of
    PBLAS PDGEMM and ABFT PDGEMM with 0 failure.  On the left,
    performance for PBLAS PDGEMM (plain) and performance for ABFT
    PDGEMM (dashed).  On the right, overhead of ABFT PDGEMM with
    respect to PBLAS PDGEMM.  }
\end{figure}

In Figure~\ref{fig:perf04}, we present a strong scalability
experiment.  The right graph assesses our claim. We see that for a
fixed problem size, when we increase the number of processors, the
overhead of the fault tolerance decreases to 0. We also observe that
the problem size is not relevant in term of overhead, the overhead is
only governed by the number of processors. We do not know of any other
fault tolerant scheme that possesses these two qualities.

\bibliographystyle{plain}
\bibliography{bddl08}

\end{document}